\documentclass[a4paper,twoside]{article}
\newcommand{\affil}[1]{$^{\rm #1}$}
\usepackage[authoryear]{natbib}
\bibpunct{(}{)}{;}{a}{}{,}
\usepackage{graphicx}
\date{} 
\title{\large\bf\flushleft The Sydney University Stellar Interferometer - A Major
Upgrade to Spectral Coverage and Performance}
\author{\parbox{\textwidth}{\flushleft
\vspace{-0.5cm}
{\it J.~Davis\affil{A,D}, M.~J.~Ireland\affil{B},
J.~Chow\affil{A}, A.~P.~Jacob\affil{A}, R.~E.~Lucas\affil{A},
J.~R.~North\affil{A}, J.~W.~O'Byrne\affil{A},
S.~M.~Owens\affil{A}, J.~G.~Robertson\affil{A},
E.~B.~Seneta\affil{C},
W.~J.~Tango\affil{A} and P.~G.~Tuthill\affil{A}}\\
\vspace{0.4cm}
{\small \affil{A}\,School of Physics, University of Sydney, NSW 2006, Australia}\\
{\small \affil{B}\,Planetary Science, MS 150-21, Caltech, 1200 E. California Blvd, Pasadena, CA 91125, USA}\\
{\small \affil{C}\,Astrophysics Group, Cavendish Laboratory, Cambridge University, Cambridge, CB3 0HE, UK}\\
{\small \affil{D}\,Email: j.davis@physics.usyd.edu.au}}}
\begin{document}
\twocolumn[
\begin{changemargin}{.8cm}{.5cm}
\begin{minipage}{.9\textwidth}
\vspace{-1cm}
\maketitle
%
%
\small{\bf Abstract:} A new beam-combination and detection system
has been installed in the Sydney University Stellar Interferometer
working at the red end of the visual spectrum
($\lambda\lambda$500--950$\,$nm) to complement the existing
blue-sensitive system ($\lambda\lambda$430--520$\,$nm) and to
provide an increase in sensitivity.  Dichroic beam-splitters have
been introduced to allow simultaneous observations with both
spectral systems, albeit with some restriction on the spectral
range of the longer wavelength system
($\lambda\lambda$550--760$\,$nm).  The blue system has been
upgraded to allow remote selection of wavelength and spectral
bandpass, and to enable simultaneous operation with the red system
with the latter providing fringe-envelope tracking.  The new
system and upgrades are described and examples of commissioning
tests presented.  As an illustration of the improvement in
performance the measurement of the angular diameter of the
southern F supergiant $\delta$~CMa is described and compared with
previous determinations.

\medskip
\medskip
\end{minipage}
\end{changemargin}
]
\small

\section{Introduction}\label{sec:intro}

The Sydney University Stellar Interferometer (SUSI)
\citep{99susi1} is a long baseline optical interferometer located
at the Paul Wild Observatory of the Commonwealth Scientific and
Industrial Research Organisation (CSIRO).  The observatory is
approximately 400\,km NNW of Sydney at latitude
-30$^{\circ}$19$^{\prime}$ and longitude
149$^{\circ}$34$^{\prime}$ East.  SUSI was designed to have two
beam-combining systems, one operating in the blue part of the
spectrum and the second optimised for the red. In its initial
configuration only the blue sensitive system was implemented. This
was designed to operate in the spectral range
$\lambda\lambda$430--520\,nm and was restricted to using narrow
spectral bandwidths ($\leq$4\,nm) with a limiting magnitude of
$B\sim+2.5$.

The provision of two separate optical tables meant that a new
beam-combination and detection system could be developed and
installed without interrupting the observing programme of bright
early-type binary stars being carried out with the original blue
system.  The primary motivation for the development of the new
system was to significantly improve the sensitivity by using a
different detection technique, more sensitive detectors, and to
take advantage of the reduced effects of atmospheric turbulence at
longer wavelengths. Therefore the new beam-combination system has
been designed for the red end of the visual spectrum
($\lambda\lambda$500--950\,nm).  The system, for which a brief
outline has been given by \citet{04spie}, has been commissioned
and is now being employed in a multi-faceted observing programme.
We will refer to it as the `red' system to distinguish it from the
`blue' system.  The increased sensitivity is discussed in
Section~\ref{sec:sensy}. It has significantly improved the
calibration of the visibility measurements, and hence the
determination of angular diameters, and it has enabled the
measurement of later spectral types than is possible with the blue
system.

On completion of the commissioning of the red system attention was
given to upgrading the performance of the blue system and this has
been described briefly by \citet{06spie}.  Dichroic beam-splitters
have been introduced to split the incoming starlight between the
red and blue systems.  The blue beam-combination system has been
re-organised with additional facilities to allow it to be used in
parallel with the red system with the latter providing delay
tracking for both systems.  The changes are described in more
detail in Section~\ref{sec:blue}.

\section{The Red Beam-Combination and Detection System}\label{sec:system}

\begin{figure}[h]
\begin{center}
\resizebox{\hsize}{!}{\includegraphics{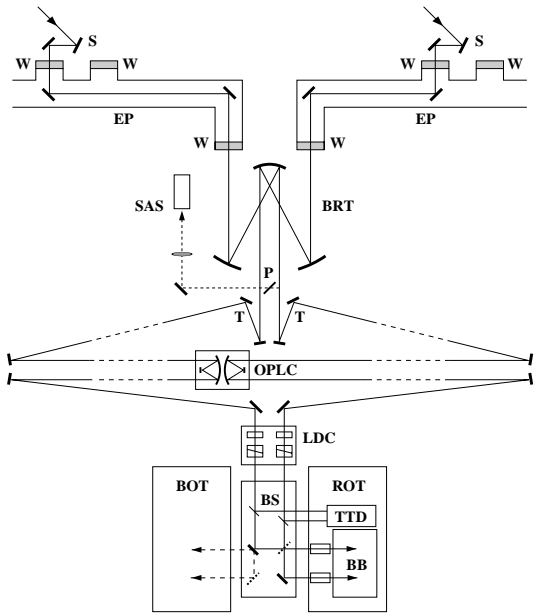}}
  \caption{The overall layout of the optics. Key: \textbf{S}
  200\,mm siderostat mirror (only two stations shown);
  \textbf{W} 180\,mm diameter window; \textbf{EP} evacuated pipe;
  \textbf{BRT} beam-reducing telescope; \textbf{P} pellicle;
  \textbf{SAS} star acquisition system; \textbf{T} wavefront tip-tilt correcting mirror;
  \textbf{OPLC} optical path-length compensation carriage;
  \textbf{LDC} longitudinal dispersion corrector; \textbf{BS} beam-splitters;
  \textbf{BOT} `blue' optical table; \textbf{ROT} `red' optical table;
  \textbf{BB} optical bread-board for `red' beam-combination optics;
  \textbf{TTD} wavefront tip-tilt detector (see Fig.~\ref{fig:ttdet} for more detail).
  Further details are given in the text.}
  \label{fig:layout}
\end{center}
\end{figure}

The overall layout of the SUSI optical system is shown
schematically in Figure~\ref{fig:layout}.  The layout is
essentially the same as that shown in Figure~2 of \citet{99susi1}
with some small but significant changes.  An additional reflection
has been introduced in the south beam prior to the beam-reducing
telescope (not shown in the figure) to make the polarisation
properties of the two arms symmetrical and some changes have been
made to the layout of the optical path length compensator (OPLC)
to reduce the number of reflections.  The two beams of starlight
from the north and south arms of the baseline array, after passing
through the OPLC and the longitudinal dispersion corrector (LDC),
are relayed to the platform between the two optical tables as
shown in Figure~\ref{fig:layout}. We note that the LDC, which was
described by \citet{99susi1}, is operated in `single-glass' mode
with only the BK7 glass as the compensating medium when used with
the red system.  This is possible because the dispersion is
significantly less in the red than in the blue region of the
spectrum.  The most significant change from Figure~2 of
\citet{99susi1} is the inclusion of the second optical table which
illustrates the relative locations of the red and blue
beam-combination systems.  Initially, starlight was directed to
either the blue or red table by manually changing the mirrors on
the central pier as shown in Figure~\ref{fig:layout}.

\begin{figure}[h]
\begin{center}
\resizebox{\hsize}{!}{\includegraphics{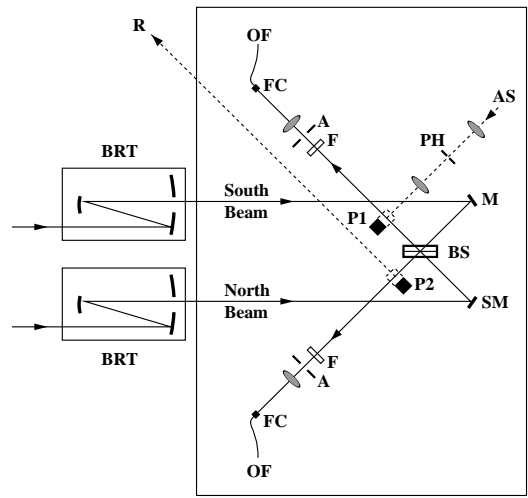}}
  \caption{A schematic of the layout of the red system optics.
  The large rectangle is the outline of the optical breadboard
  shown in Fig.~\ref{fig:layout}.
  Key:\textbf{BRT} beam-reducing telescope; \textbf{M} fixed mirror;
  \textbf{SM} scanning mirror; \textbf{BS} beamsplitter;
  \textbf{P1}and \textbf{P2} slide-mounted pentaprisms;
  \textbf{F} filter; \textbf{A} aperture; \textbf{FC} optical fibre clutch;
  \textbf{OF} optical fibre; \textbf{PH} pinhole; \textbf{AS} artificial `star'
  light source; \textbf{R} reference beam.  Further details are given in the text.}
  \label{fig:redot}
\end{center}
\end{figure}

The layout of the optics for the red system basically follows that
for the blue system \citep{99susi1} and is shown schematically in
Figure~\ref{fig:redot}.  There are, however, a few key
differences. These include a new wavefront tip-tilt detection
system based on a CCD camera, additional beam reducing telescopes
at the input to the beam-combination system to allow for a more
compact system and to provide a location for field stops or
spatial filters, wide-band spectral filters combined with a fringe
envelope scanning mirror to increase the sensitivity, and
avalanche photodiode (APD) detectors in place of photomultipliers
to give high detected quantum efficiency in the red region of the
spectrum.  These key changes are discussed in the following
sections.  The introduction of dichroic beam-splitters and the
re-arrangement of the blue system is discussed in
Section~\ref{sec:blue}.

\subsection{The Optical System}\label{sec:red}

The two beams of starlight from the north and south arms, on
reaching the central pier, are first divided by partially
polarising beam-splitters (mean reflectance at $45\deg$ for
$\lambda\lambda$440--900\,nm equals 0.64 for the s-polarisation
and 0.36 for the p-polarisation giving a mean reflectance of 0.5).
The reflected beams are focussed by a Celestron 8 telescope onto
the chip of a CCD camera mounted on the red optical table.
Apertures in front of the telescope define the beam size. The
images are used to detect atmospheric turbulence induced wavefront
tilts and to provide feedback to the tip-tilt mirrors (see
Section~\ref{sec:wobble}). The transmitted beams pass through
apertures to define the size of the signal beams.  Aperture
diameters of 28\,mm are generally employed (equivalent to 84\,mm
on the sky).  The beams are then reflected by mirrors onto the red
table where the beam diameters are reduced by individual
beam-reducing telescopes.  Each of the beam-reducing telescopes
consists of two concave paraboloidal mirrors arranged in a
confocal configuration to produce a collimated output.  The
primary mirrors have focal lengths of 405\,mm and the secondaries
focal lengths of 98.5\,mm giving a reduction in the beam diameter
by a factor of 4.1. The primary mirror of each telescope has a
hole trepanned in its centre.  The input beam is incident on the
primary mirror to one side of the central hole and the output beam
reflected from the secondary mirror passes through the central
hole in the primary. Pinholes can be placed at the common foci of
the paraboloidal mirrors to act as field stops for isolating
individual stars of visual binaries or for spatial filtering.  The
beams exiting the beam-reducing telescopes enter a
beam-combination system with the layout shown in schematic form in
Figure~\ref{fig:redot}. Mirrors reflect the incoming beams into a
partially reflecting dielectric beam-splitter at $45^{\circ}$. The
reflectivity of the beam-splitter is within 10 per cent of 0.50
over the spectral range 500--900\,nm at the design angle of
$45^{\circ}$ for the polarisation with the E-field horizontal.  A
major difference from the blue system is that the mirror
reflecting the north beam into the beam-splitter is mounted on a
piezo-actuated linear translator that allows the mirror to be
scanned through the fringe envelope. The combined beams exiting on
the two sides of the beam-splitter pass through matched broad-band
interference filters before being focussed as ~f/12 beams on
100\,$\mu$m diameter multimode optical fibre feeds to APD
detectors located on the roof of the instrument enclosure. There
is provision for the insertion of wire grating polarisers prior to
the focussing lenses. The matched interference filters are mounted
in filter wheels so that it is possible to switch rapidly between
filters with different central wavelengths and spectral
bandwidths. Several pairs of matched broad-band filters (with
spectral bandwidths of $\sim$5\% and $\sim$10\%) are available but
most observations to date, including those presented here, have
been made using filters centred on 700\,nm with a spectral
bandwidth of 80\,nm.

\subsubsection{Alignment aids} \label{sec:align}

The critical alignment of a long-baseline interferometer and the
alignment aids and techniques employed in SUSI for the blue
beam-combining optics have been discussed by \citet{99susi1}.
Similar additional optics and an `artificial star', identified in
Figure~\ref{fig:redot} by \textbf{AS}, have been included for the
red table.  The artificial star is a diffraction--limited pinhole
that can be illuminated with a 633\,nm HeNe laser for alignment
purposes or by a Maglite torch bulb (manufactured by Mag
Instruments Inc.) that acts as a white light source for
establishing matched path positions through the white light
fringe. Pentaprisms are used instead of the periscope rhombs used
on the blue table \citep{99susi1} and these are mounted on
precision slides driven by linear stepping motors with Hall effect
position sensors. They are used to insert the laser or white light
beams into the beam-splitter along the optical axis of the
interferometer or to relay a reference beam of incoming starlight
to the CCD based tip-tilt detection system.  An additional
pentaprism, not shown in Figure~\ref{fig:redot}, is used to
reflect light to a CCD video camera to assist with alignment.

Two improvements have been made that do not affect the sensitivity
of the instrument but which have facilitated alignment and
improved the reliability and efficiency of operation.  They are
the installation of a new local area network and the introduction
of a large number of New Focus picomotor drives in place of
micrometers for remote mirror and beam-splitter control. The
picomotors are operated through the network from a hand-terminal
that can be patched to any location within the instrument and this
enables alignment to be performed efficiently by one person.  The
crucial beam-splitters that align the tip-tilt images relative to
the reference images to ensure accurate overlap of the science
beams are equipped with picomotors and, as well as being
adjustable via the hand terminal, they are computer-controlled to
permit automatic alignment with starlight during observing
sessions.

\subsection{The Star-Acquisition System}\label{sec:acqu}

The SUSI star-acquisition system has been described by
\citet{99susi1} and in detail by \citet{91ebs}.  A pellicle
beamsplitter can be moved into either the north or south beam to
reflect $\sim$8 per cent of the incident light to an intensified
CCD camera. Originally the camera fed a video signal to custom
hardware that provided the logic for determining the pointing
error in real time.  The hardware was interfaced with a
single-board computer and an MS-DOS based PC to provide feedback
signals to correct siderostat pointing during star acquisition. To
overcome problems with the now obsolete video digitiser chip the
system has been upgraded and the video signal now goes directly to
a `framegrabber' card installed in a PC running Microsoft Windows.
The logic is now integrated into new software that displays the
video signal and links the acquisition system directly with the
new tip-tilt servo.

\subsection{Tip-Tilt Correction}\label{sec:wobble}

\begin{figure}[h]
\begin{center}
\resizebox{\hsize}{!}{\includegraphics{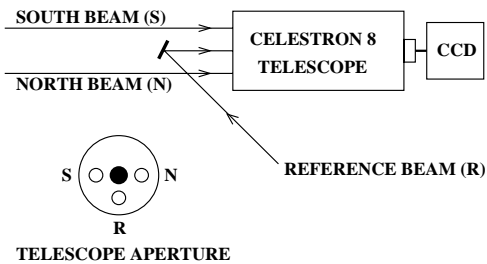}}
  \caption{Schematic layout of the wavefront tip-tilt detection system.
   The relative positions of the input beams on the aperture of
   the telescope are shown in the lower diagram.  Further details
   are given in the text.}
  \label{fig:ttdet}
\end{center}
\end{figure}

The main components of the new tip-tilt system are the CCD
detector shown in Figure~\ref{fig:ttdet} and the tip-tilt mirrors
labelled \textbf{T} in Figure~\ref{fig:layout}.  The tip-tilt
mirrors are activated by two-axis piezo-electric actuators.  Two
Astromed/Astrocam cameras have been used as the detector: the
original one with an EEV 30-A CCD chip and a newer one with a
back-illuminated SITe 502A chip. The overall control of the system
is managed by a PC running real-time Linux.

The CCD accepts broad-band light which, in practice, means that
wavelengths between 400 and 950\,nm are used for tip/tilt
detection.  The broad optical bandwidth is employed so that
adequate photon count rates can be obtained when observing
relatively faint stars.   There are no atmospheric refraction
correctors so the width and shape of the point-spread functions
used for tip-tilt tracking are a function of seeing, stellar
spectral type, and the zenith-distance of the source.

\begin{figure}[h]
\begin{center}
\includegraphics[scale=0.36]{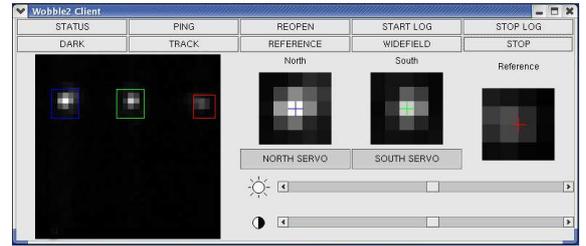}
 \caption{The graphical user interface for the new tip/tilt system,
 showing the images for the servos locked on a 633\,nm laser
 signal autocollimated from the siderostats.}
 \label{fig:ttgui}
\end{center}
\end{figure}

The configuration of the three image readout regions in the corner
of the CCD nearest to its readout region is shown in
Figure~\ref{fig:ttgui}.  In this standard observing mode the
pixels are square with each side equivalent to 0.81\,arcsec on the
sky. The 5\,pixel by 5\,pixel readout regions for the north and
south beams enable the stellar point spread functions to be
sampled. This in turn enables centroid estimation that is less
dependent on the shape of the point-spread function than for a
quad-cell centroid estimator. All images and tip-tilt mirror
positions are saved during observing and the data sets are used to
estimate residual seeing effects in post-observational processing.
A 20\,ms sample time is generally used for observing in this
configuration, although a 10\,ms sample time is available for the
(rarely used) quad-cell 2\,pixel by 2\,pixel binning mode. These
relatively long sample times seriously limit the tip-tilt
performance in fast seeing conditions, but are necessary because
of long CCD readout times and the requirement for similar tip-tilt
performance on both bright and fainter stars.  A new camera
incorporating an electron-multiplying CCD (EMCCD) is being
commissioned to give improved signal-to-noise and faster read-out.

Fine alignment of the tip-tilt system is achieved using the
reference beam labelled \textbf{R} in Figures~\ref{fig:redot} and
\ref{fig:ttdet}. This beam travels through a 4:1 beam expander
before being imaged by the Celestron~8 telescope onto the 4\,pixel
by 4\,pixel reference readout region shown in
Figure~\ref{fig:ttgui}. While the tip-tilt servo is locked, the
north and south beams are viewed one at a time in the reference
region by closing and opening appropriate shutters on the red
table.  Picomotors adjust the orientation of the north and south
tip-tilt beam-splitters so that the images formed by the north and
south beams are individually centred in the reference region. This
ensures that the north and south beams are overlapped with their
wavefronts plane parallel at the beam-combiner (\textbf{BS} in
Figure~\ref{fig:redot}).  The tip-tilt servo, picomotor and red
table slide and shutter controllers are all connected to the local
area network, enabling alignment of the tip-tilt system to be
performed automatically.

\subsection{The Optical Path-Length Compensation (OPLC)
System}\label{sec:oplc}

The OPLC system in SUSI has been described in detail by
\citet{99susi1}.  In brief, a carriage carrying back-to-back cat's
eye retro-reflectors moves differentially between starlight from
the north and south beams over a 70\,m long track.  The position
and velocity of the carriage are monitored and controlled by the
difference signal between laser metrology systems mounted at each
end of the track.  A number of improvements have been made to the
OPLC system.  These include:

\begin{itemize}
    \item The system has been ported and streamlined from the original, now obsolete,
    Solaris workstation and VME crate to a PC running Linux.
    \item The control and user interface software has been
    upgraded and communication has been established with the fringe
    detection system to allow feedback of the current fringe
    envelope centre position and the servoing of the carriage
    position (Section~\ref{sec:detect}).
    \item The original long and short low-voltage piezo-electric actuators
    (PEAs) on which the small flat mirrors at the foci of the cat's eye
    retro-reflectors were mounted have been replaced by two
    nominally identical PEAs.  The new PEAs act in a push-pull
    arrangement, each compensating half the tracking-error signal,
    with an increased bandwidth of 500\,Hz.  Each of the PEAs has
    a near identical, experimentally determined range of 78\,$\mu$m in
    optical path with a step resolution of 19\,nm in optical path.
    \item The fact that the two PEAs are nominally identical, and
    have been shown to have similar expansion and hysteresis
    characteristics, allows operation with a single metrology laser
    operating from one end of the OPLC.  This facility has been
    commissioned to provide a back-up mode in the event of a
    metrology laser failure.
    \item The stepper motor that drives the carriage along
    the rails has been replaced with a new stepper motor.  The new motor shows
    lower frequency and significantly reduced RMS deviations from linearity
    and it follows that these are more efficiently corrected by the piezo-mounted secondary
    mirrors of the cat's eyes.
    \item The high-friction, high-rigidity tyre on the drive wheel
    that is directly coupled to the stepper motor has been
    replaced.  It was found to have worn unevenly resulting in large systematic
    optical path tracking errors.  Steel wheels of the same outside diameter
    as the tyre have been installed and improved tracking has been achieved.
    \item Non-linearities in the motor drive and in the hysteresis
    of the PEAs have been modelled and used to optimise the
    control software.
\end{itemize}

\subsection{Fringe Detection and Data Acquisition}\label{sec:detect}

\begin{figure}[h]
\begin{center}
\includegraphics[scale=0.47]{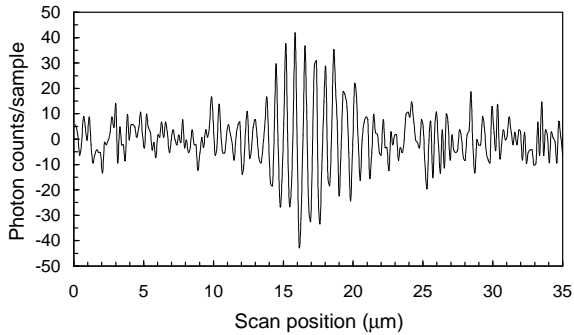}
\caption{An example of a single short raw fringe scan for
$\delta$\,CMa at a baseline of 5m on 21 April, 2004.  The photon
counts/sample time are relative to the mean count for the scan and
the scan consists of 256 samples each 0.2\,ms long.   The mean
count per sample for the scan was 43.} \label{fig:fringe}
\end{center}
\end{figure}

\begin{figure}[h]
\begin{center}
 \includegraphics[scale=0.50]{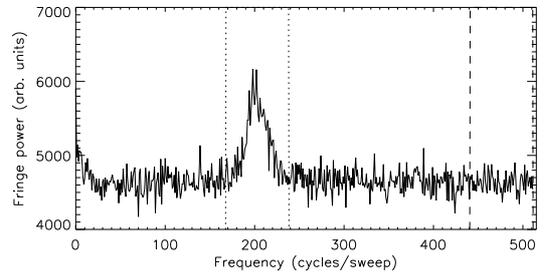}
 \caption[An integrated power spectrum for observations of HR~5264]
{An integrated power spectrum of $\tau$~Vir (HR~5264) (\textit{R}
magnitude = 4.2) for 1000 fringe scans observed at a baseline of
80\,m  in 1.8\,arcsec seeing.  The dotted lines show the edges of
the rectangular window function (see text), and the dashed line
shows the equivalent window for estimating bias. Note that the
y-ordinate reaches zero  well below the x-axis, demonstrating that
the fringe signal centered around 200 cycles/sweep is dominated by
bias.}
 \label{fig:power}
\end{center}
\end{figure}

The fringe detection system consists of two Perkin Elmer Photon
Counting Module APD detectors (fed by the optical fibres shown in
Figure~\ref{fig:redot}) and a mirror (labelled SM in
Figure~\ref{fig:redot}) mounted on a piezo-electric actuator that
scans through an optical path up to 140\,$\mu$m long.  The
actuator is driven to and fro by a triangular waveform generated
from lookup tables to eliminate non-linearities in the motion. The
fringe signal is the difference in count rates between the two
APDs. This is similar to some features of the system developed for
COAST \citep{coast}.  An example of a fringe scan using a short
scan of length 35\,$\mu$m is shown in Figure~\ref{fig:fringe}. The
control software for the fringe detection and data acquisition
system runs on a real-time Linux PC. As well as the basic tasks of
controlling the scanning mirror and saving the raw data streams,
the control computer calculates the fringe power spectrum (shown
in Figure~\ref{fig:power}), estimates the squared fringe
visibility, the signal-to-noise ratio, and the fringe position.

In the low signal-to-noise regime, finding the fringe position is
aided by forming the envelope of fringe power. This fringe
envelope is the squared modulus of the analytic signal as defined
by \cite{65rb}. It is calculated by windowing the fringe peak in
the positive half of the fringe Fourier transform, forming the
inverse transform and taking its squared modulus.  The upper and
lower bounds of the rectangular window function are shown by the
dotted lines in Figure~\ref{fig:power}.  A fading memory average
of the fringe power spectrum and fringe envelope function are
displayed to the operator in a graphical user interface, and the
first moment of the bias-subtracted fringe envelope is used for
tracking the fringes. Details of the fringe tracking algorithm are
discussed by \citet{mjithesis}.

\section{Observing with the Red System}\label{sec:observing}

The observing procedure has been automated with a software
scheduler that controls the sequence of operations that constitute
a complete observation.  This not only eases the load on an
observer but also ensures that no steps in the observing procedure
are missed.  Each stellar observation includes a set of fringe
scans and photometric and background scans.  A complete
observation of a star including acquisition, fringe scans and
photometric scans is completed in an average of 6\,min. Each
observation of a programme star is generally interleaved between
observations of `calibrator' stars so that a sequence of
`calibrator-programme star-calibrator' is completed in
$\sim$18\,min. Calibration is discussed further in
Section~\ref{sec:cal}.

\subsection{The Scheduler}\label{sec:sched}

The scheduler software accepts the HR number of the star to be
observed and causes an astrometric file with position and tracking
velocities for the siderostats and OPLC to be created.  It then
initiates a sequence of operations.  The siderostats and OPLC
carriage are driven to their predicted positions and set moving at
their tracking velocities. The north and south images from the
siderostats are located successively by the star-acquisition
system which provides correction signals to the siderostats to
bring the images within the field of view of the tip-tilt system.
At this point the tip-tilt system takes over and closes the
feedback servo to the siderostats. With the tip-tilt servo on, the
OPLC is scanned through a preset range to find and acquire
fringes.  A preset number of fringe scans are then recorded and,
on completion of the scans, shutters are moved in and out of the
beams to obtain photometric measurements for each beam.  The
siderostats are stopped and the background is measured with the
star out of the field of view. The scheduler then accepts the next
HR number from a command line input or from a scheduler file.  The
observer is alerted to any errors or problems in the sequence of
operations and is able to interrupt the scheduler at any time and
subsequently resume where it stopped.  Prior to the commencement
of an observing session, and approximately every hour during the
session, the scheduler is instructed to initiate a fine alignment
of the tip-tilt system.

\subsection{Calibration}\label{sec:cal}

As described in Sections~\ref{sec:detect} and \ref{sec:analysis},
SUSI measures the square of the interference fringe visibility
($V^{2}$). In practice this will be less than the true $V^{2}$ due
to instrumental and residual seeing losses.  The magnitude of
these losses is established by interleaving the observations of a
programme star between observations of `calibrators'. A calibrator
is ideally a star whose true $V^{2}$ can be accurately
predicted---either because it has a small angular size compared to
the programme star or because its angular diameter is known
accurately.

There are a number of desirable characteristics for a calibrator
and these include being single, non-variable, and close to the
programme star in the sky. Ideally a calibrator would be of
similar magnitude and spectral type as the programme star but this
implies that the stars are nearly identical and it follows that
this does not make for a suitable calibrator. In practice,
compromises have to be made and these depend heavily on the
limiting magnitude of the instrument. The problem experienced in
calibrating observations for the blue system was the fact that the
relatively bright limiting magnitude made it very difficult to
find suitable calibrators close to a programme star.  The increase
in sensitivity of the red system has eased this problem.

The calibration procedure essentially involves defining a transfer
function $T$ for a calibrator by

\begin{equation}
T = \frac{V^{2}_{\mathrm{obs}}}{V^{2}_{\mathrm{exp}}}
\label{eqn:tf}
\end{equation}
where $V^{2}_{\mathrm{obs}}$ is the observed $V^{2}$ and
$V^{2}_{\mathrm{exp}}$ is the expected $V^{2}$ for the calibrator.
If $V^{2}_{\mathrm{obs*}}$ is the observed $V^{2}$ for the
programme star, its true visibility squared
$V^{2}_{\mathrm{true*}}$ is given by

\begin{equation}
V^{2}_{\mathrm{true*}} =
\frac{V^{2}_{\mathrm{obs*}}}{\overline{T}} \label{eqn:tftrue}
\end{equation}
where $\overline{T}$ is the mean of the transfer functions of
calibrators observed either side of the programme star.

\subsection{The Effective Wavelength}\label{sec:efflam}

The effective wavelength of observations made with the red system
will be a function of the spectral type of the star being observed
and also the filters employed because of their wide spectral
bandwidths. The effective wavelength $\lambda_{\mathrm{eff}}$ is
given by

\begin{equation}
\frac{1}{\lambda_{\mathrm{eff}}} = \sigma_{\mathrm{eff}} =
\frac{\int_{0}^{\infty}I^{2}(\sigma)\sigma\mathrm{d}\sigma}{\int_{0}^{\infty}I^{2}(\sigma)\mathrm{d}\sigma}
\label{eqn:efflam1}
\end{equation}
with
\begin{equation}
I^{2}(\sigma) = T(\sigma)N^{2}(\sigma)S^{2}(\sigma)
\end{equation}
where $\sigma$ is the wavenumber equal to $1/\lambda$, $T(\sigma)$
is the transfer function (see equation~(\ref{eqn:tf})),
$N(\sigma)$ is the photon flux from the star per unit wavenumber
interval, and $S(\sigma)$ is the spectral response of the
instrument.

The effective wavelength as a function of intrinsic
(\textit{B}-\textit{V})$_{\mathrm{o}}$ for main sequence stars has
been evaluated for the 700\,nm filters via computation of the
spectral response of the system and also by measurements.  The
spectral response $S(\sigma)$ has been computed taking into
account the spectral transmission of the optical system, including
window transmission, mirror reflectance, beam-splitter
transmission/reflectance, transmission of the matched interference
filters, the quantum efficiencies of the APDs as a function of
wavenumber, and the transmission of the atmosphere.  The transfer
function includes the loss in $V^{2}$ due to residual seeing
effects and an approximate allowance for this has been included in
the computations.  The losses in $V^{2}$ for no correction and for
100\% correction of wavefront tip-tilt have been published by
\citet{tandt}. The losses for 1.5\,arcsec seeing and $\sim$50\%
correction have been derived as a function of wavenumber from the
Tango \& Twiss results as representative of average conditions for
SUSI observations and used in the calculation of effective
wavelength. It is noted that the extremes of 1\,arcsec seeing with
full tip-tilt correction and 2\,arcsec seeing with no correction
only change the effective wavelength by $\pm$0.06\%. Effective
wavelengths were computed for spectral types O5\,V, B6\,V, A2\,V,
A8\,V, F3\,IV, G2\,IV, G8\,IV, K4\,V and K5\,V using photon flux
distributions derived from the library of stellar spectra by
\citet{92silva}.

Measurements of the effective wavelength have also been made from
sets of scans recorded for selected main-sequence stars.  The scan
steps were calibrated by means of scans taken in autocollimation
with the 633\,nm HeNe laser illuminating the `artificial star'
pinhole (Section~\ref{sec:align}).  In the case of the stellar
measurements, the power spectrum of each sweep, determined in the
data-processing pipeline to be described in
Section~\ref{sec:pipe}, is an estimate of $I^{2}|\gamma|^{2}$
where$|\gamma|$ is the degree of coherence at the observing
baseline.  In the pipeline the power spectra are averaged and are
used to estimate the fringe visibility.  The first moment of the
power spectrum is

\begin{equation}
\sigma_{1} =
\frac{\int_{0}^{\infty}I^{2}(\sigma)|\gamma(\sigma)|^{2}\sigma\mathrm{d}\sigma}{\int_{0}^{\infty}I^{2}(\sigma)|\gamma(\sigma)|^{2}\mathrm{d}\sigma}
\label{eqn:efflam1}
\end{equation}

For observations made where $|\gamma(\sigma)|^{2} \sim 1$, we have
$\sigma_{1} = \sigma_{\mathrm{eff}}$.  The main-sequence stars
selected for the measurement of $\sigma_{\mathrm{eff}}$, and hence
$\lambda_{\mathrm{eff}}$, were observed at a baseline of 5\,m such
that this condition was met. The stars and the effective
wavelengths determined for them are listed in
Table~\ref{tab:meas_efflam}.

\begin{table}
\begin{center}
  \caption[Effective wavelength]{The measured values of effective
  wavelength($\lambda_{\mathrm{eff}}$)
  for selected main-sequence stars measured at a 5\,m baseline.  N is the number
  of measurements and SEM the standard error in the mean value of effective wavelength.}
  \label{tab:meas_efflam}
  \begin{tabular}{ccccc}
\hline
Star &  $(B-V)_{0}$ & N &  $\lambda_{\mathrm{eff}}$ & SEM \\
     &             &   &  (nm) & (nm) \\
 \hline
$\tau$ Sco &  -0.30 & 18 & 695.20 & 0.17 \\
$\kappa$\,Vel & -0.24 & 5 & 695.93 & 0.48 \\
$\sigma$\,Leo &  -0.05 & 4 & 694.97 & 0.84 \\
$\alpha$\,CMa & 0.00 & 8 & 693.69 & 0.34 \\
$\iota$\,Cen &  0.04 & 10 & 695.33 & 0.43 \\
$\phi$\,Leo &  0.20 & 4 & 694.83 & 1.27 \\
$\beta$\,TrA &  0.30 & 10 & 695.33 & 0.22 \\
$\beta$\,Vir &  0.52 & 7 & 697.26 & 0.87 \\
$\beta$\,Vir &  0.52 & 6 & 695.90 & 0.87 \\
\hline
\end{tabular}
\end{center}
\end{table}

\begin{figure}[h]
\begin{center}
\includegraphics[scale=0.55]{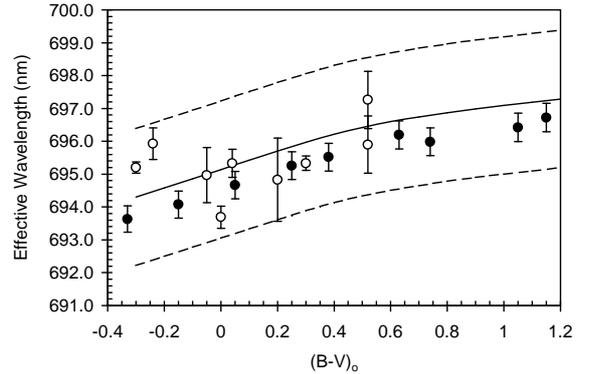}
\caption{The computed and measured values of effective wavelength
with the adopted relationship shown by the solid line.  Key:
$\bullet$ - computed values; $\circ$ - measured values.  The
estimated uncertainty in the adopted relationship of $\pm$0.3\% is
represented by the two dashed lines.} \label{fig:efflam}
\end{center}
\end{figure}

The computed and measured values of $\lambda_{\mathrm{eff}}$ are
plotted against $(B-V)_{0}$ in Figure~\ref{fig:efflam}. There is
excellent agreement between the measured and calculated values
with the mean difference equal to $\sim0.08$\%.  The shape of a
smooth curve drawn through the calculated points is insensitive to
changes in the parameters that enter the calculations and only the
vertical positions of the calculated values in
Figure~\ref{fig:efflam} are affected.  The adopted scale of
effective wavelength for main-sequence stars is therefore based on
a smooth curve drawn through the calculated points shifted by
$+$0.6\,nm ($\sim0.08$\%) to place it close to the mean difference
between the measured and calculated values and this is represented
by the solid line in Figure~\ref{fig:efflam}.  A conservative
uncertainty of $\pm$0.3\% ($\sim\pm$2\,nm) has been adopted and
this is shown by the dashed lines in Figure~\ref{fig:efflam}.  The
effective wavelengths corresponding to the solid line in the
figure are tabulated against $(B-V)_{0}$ in
Table~\ref{tab:efflam}.

\begin{table}[h]
\begin{center}
  \caption[Effective wavelength]{The adopted scale of effective wavelength
  $\lambda_{\mathrm{eff}}$ v. (\textit{B}-\textit{V})$_{\mathrm{o}}$ for
  SUSI's 700\,nm system for non-reddened main-sequence stars.  Details are given
  in the text.}
  \label{tab:efflam}
  \begin{tabular}{cccc}
\hline (\textit{B}-\textit{V})$_{\mathrm{o}}$ & $\lambda_{\mathrm{eff}}$ &  (\textit{B}-\textit{V})$_{\mathrm{o}}$ & $\lambda_{\mathrm{eff}}$  \\
 & (nm) & & (nm) \\
\hline
-0.30 & 694.3 & 0.50 & 696.4 \\
-0.20 & 694.6 & 0.60 & 696.6 \\
-0.10 & 694.9 & 0.70 & 696.7 \\
0.00 & 695.1 & 0.80 & 696.9 \\
0.10 & 695.4 & 0.90 & 697.0 \\
0.20 & 695.7 & 1.00 & 697.1 \\
0.30 & 696.0 & 1.10 & 697.2 \\
0.40 & 696.2 & 1.20 & 697.3 \\
\hline
\end{tabular}
\end{center}
\end{table}

The effective wavelength for reddened main-sequence stars or stars
of other luminosity classes, reddened or not, can be computed from
the observed stellar flux distribution corrected for atmospheric
extinction following the procedure described here.

\section{Data Analysis}\label{sec:analysis}

A software pipeline has been developed to process the recorded
fringe scans and this is described in the following section.

\subsection{Data Pipeline}\label{sec:pipe}

\begin{figure}[h]
\begin{center}
 \includegraphics[scale=0.55]{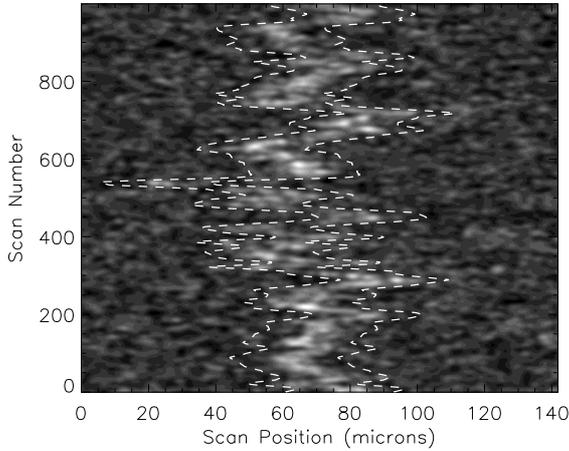}
 \caption[Smoothed fringe envelopes for observations of HR~5264]
{The smoothed fringe envelopes (see Section~\ref{sec:pipe}) for a
set of 1000 scans for 80\,m baseline data on $\tau$~Vir (HR~5264)
(\textit{R} magnitude = 4.2) in 1.8\,arcsec seeing.  Dashed lines
join the edges of the spatial-domain window. Signal-to-noise per
scan is 0.55 before spatial-domain windowing and 0.8 after.  The
integrated power spectrum for these data is shown in
Figure~\ref{fig:power}.}
 \label{fig:posest}
 \end{center}
\end{figure}

The data analysis pipeline first forms the power spectrum of the
fringes and the fringe envelope. Once the fringe envelope has been
calculated for each scan, a moving spatial-domain window function
is defined as shown in Figure~\ref{fig:posest}. This window is
based on an estimate for the fringe position formed by looking for
the peak in the set of fringe envelopes smoothed in space and
time. Only the data within this spatial domain window are
considered in forming the final estimate for the observed
$V^{2}_{\mathrm{obs}}$. The details of the determination of
$V^{2}_{\mathrm{obs}}$, which is proportional to the sum of fringe
power (minus bias) normalised by the product of the mean APD
fluxes, are given by \citet{mjithesis}.

Two important quantities are extracted from the recorded tip-tilt
data. The first, $\sigma_t^2$, is the variance in tip/tilt system
centroid, corrected for effects of photon and readout noise. The
second, $\theta_T$ is the full-width half maximum of a Gaussian
fitted to the sum of the tip/tilt mirror position and the image
centroid. The variance in the normalised APD count rates
$\sigma_I^2$ is also used as a measure of scintillation. Together,
these three quantities are used to correct the value of
$V^{2}_{\mathrm{obs}}$ for each data set:

\begin{equation}
 V^{2}_{\mathrm{obs\_corr}} = V^{2}_{\mathrm{obs}}{\exp}(a\sigma_t^2 + b\theta_T^2 + c\sigma_I^2)
\label{eqn:seemod}
\end{equation}

\begin{figure}
\begin{center}
\includegraphics[scale=0.45]{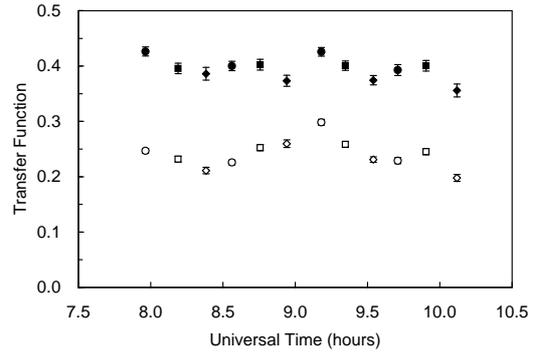}
\caption{An example of output from the data processing pipeline
for three calibrators observed at 700\,nm with a baseline of 20\,m
on 2 July 2004. Key: round symbols are for HR~3685 ($\beta$~Car),
square symbols for HR~3699 ($\iota$~Car) and diamond symbols for
HR~4199 ($\theta$~Car).  Open symbols represent transfer function
$T$ values obtained using equation~(\ref{eqn:tf}) and the filled
symbols are for the same data but with the seeing correction of
equation~(\ref{eqn:seemod}) applied.} \label{fig:cal}
\end{center}
\end{figure}

Here $V^{2}_{\mathrm{obs\_corr}}$ is the value of
$V^{2}_{\mathrm{obs}}$ corrected for residual seeing, and the
parameters $a$, $b$ and $c$ have been estimated empirically based
on fits to large data sets of calibrator observations. Further
details of this seeing correction can be found in
\citet{mjithesis, 06mji}.  The procedure has produced a
significant improvement in the calibration of the observational
data.  An example of data processed with and without the seeing
correction for three calibrators observed on 2 July, 2004, a night
of variable seeing, is shown in Figure~\ref{fig:cal}. The overall
scatter in the transfer function about the mean value has been
reduced from $\pm21$\% to $\pm9$\% for this data set.  The
improvement achieved depends on the quality of the night but,
taking an average over several nights, the application of the
seeing correction reduced the scatter in the transfer function by
a factor of $\sim1.65$.

\section{The Blue System Upgrade} \label{sec:blue}

\begin{figure}[h]
\begin{center}
\includegraphics[scale=0.4, angle=0]{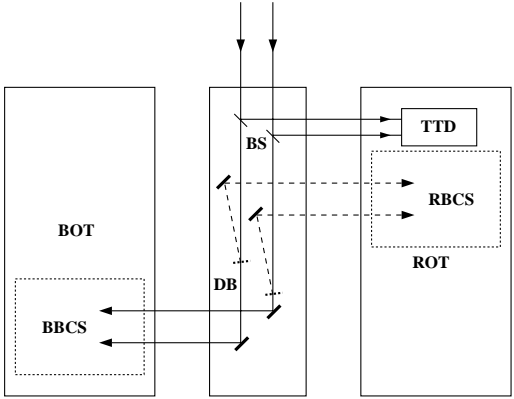}
\caption{The layout of the beam-relay system for the red and blue
beam-combining systems.  Key: BS = Beam-Splitters; TTD = Tip-Tilt
Detector; DB = Dichroic Beam-splitters; BOT = Blue Optical Table;
BBCS = Blue Beam-Combining System; ROT = Red Optical Table; RBCS =
Red Beam-Combining System.}\label{fig:dichroics}
\end{center}
\end{figure}

Following the completion of commissioning of the red system,
dichroic beam-splitters were introduced on the central pier, as
shown in Figure~\ref{fig:dichroics}, to split the incoming
starlight between the blue and red tables to enable them to be
used simultaneously.  The blue end of the spectrum, for
wavelengths up to 520\,nm, is transmitted while the red end of the
spectrum is restricted to 550--760\,nm for $>$95\% reflectance.
The dichroic beam-splitters are designed and oriented for
near-normal incidence of the incoming beams to minimize
differential polarisation effects.

The original blue system has been re-arranged to accommodate the
change in position at which the north and south beams are directed
onto the table.  The polarising beam-splitter cubes shown in
Figure~5 of \citet{99susi2} are redundant since the tip-tilt
detection is now done on the red optical table (see
Section~\ref{sec:wobble}) and they have been removed.  These cubes
had significant aberrations and their removal results in better
performance of the blue system.  The dispersing prisms have been
mounted on rotating platforms driven by stepping motors to allow
remote wavelength selection and rapid wavelength switching.  The
slits have also been fitted with stepping motors to permit remote
control of the spectral bandwidth. The red and blue systems can
now be operated simultaneously with the optical paths equalised
for the red system by fringe envelope tracking.  This will not
necessarily equalize the optical paths for the blue system and
provision has been made for small adjustments to the relative
paths of the north and south beams on the blue table. Referring to
Figure~5 of \citet{99susi2}, the mirror that directs the incoming
north beam to the beam-combiner has been mounted on a slide driven
by a stepping motor to enable small adjustments to the relative
optical paths. The original reference detector on the blue table
has been retained and enables the blue beams to be accurately
superimposed with the aid of picomotor controlled mirror
adjustments while the tip-tilt system stabilises the beams.

\section{Commissioning Results}\label{sec:results}

The corrections to the observational data for the residual effects
of seeing, as applied in the data pipeline for the red system,
have been discussed in Section~\ref{sec:pipe} and an example of
uncorrected and corrected data shown in Figure~\ref{fig:cal}. As
an example of the performance of the new red system the
determination of the angular diameter of $\delta$~CMa is presented
in Section~\ref{sec:delcma}.  The results of tests carried out on
the upgraded blue system are given in Section~\ref{sec:blue}.

\subsection{The Angular Diameter of $\delta$~CMa}\label{sec:delcma}

\begin{table*}[t]
\begin{center}
  \caption[Observations of $\delta$~CMa.]{Observations of $\delta$~CMa.
  $b$ is the mean projected baseline which varied
  by a total of $\Delta b$ during the observations.  N is the
  number of sets of scans. Each set consisted of 1000 140\,$\mu$m long
  scans, each scan consisting of 1024 0.2\,ms photon count samples.
  $V^{2}$ is the mean seeing-corrected normalised square of the fringe
  visibility.  $\sigma_{\mathrm{V^{2}}}$ is the scaled uncertainty in
  $V^{2}$ (details are given in the text).}
  \label{tab:obs}
  \begin{tabular}{rrccccl}
\hline \multicolumn{1}{c}{Date} & \multicolumn{1}{c}{$b$} &
$\Delta b$ & N &
$V^{2}$ & $\sigma_{\mathrm{V^{2}}}$ & \multicolumn{1}{c}{Calibrators} \\
 & \multicolumn{1}{c}{(m)} & (m) & & & & \\
\hline
 9 Nov 2003 &  5.00 & 0.006 & 4 & 0.980 & 0.020 & $\epsilon$~CMa, $\eta$~CMa, o$^{2}$~CMa \\
11 Nov 2003 & 19.96 & 0.004 & 2 & 0.582 & 0.026 & $\epsilon$~CMa, $\eta$~CMa, o$^{2}$~CMa \\
12 Nov 2003 & 19.94 & 0.148 & 4 & 0.598 & 0.019 & $\epsilon$~CMa, $\eta$~CMa, o$^{2}$~CMa \\
27 Feb 2004 &  4.99 & 0.007 & 3 & 0.996 & 0.022 & $\epsilon$~CMa, $\eta$~CMa, o$^{2}$~CMa \\
28 Feb 2004 &  4.99 & 0.010 & 6 & 0.968 & 0.022 & $\epsilon$~CMa, $\eta$~CMa \\
 1 Mar 2004 & 39.94 & 0.060 & 3 & 0.048 & 0.006 & $\epsilon$~CMa, $\eta$~CMa \\
30 Mar 2004 & 30.00 & 0.006 & 2 & 0.239 & 0.038 & $\epsilon$~CMa, $\eta$~CMa \\
21 Apr 2004 &  4.99 & 0.012 & 4 & 0.961 & 0.017 & $\epsilon$~CMa, $\eta$~CMa \\
 9 Jan 2005 &  5.00 & 0.010 & 8 & 0.926 & 0.018 & $\epsilon$~CMa, $\eta$~CMa \\
11 Dec 2006 & 29.96 & 0.066 & 8 & 0.230 & 0.004 & $\epsilon$~CMa, $\eta$~CMa \\
12 Dec 2006 &  9.99 & 0.023 & 6 & 0.863 & 0.024 & $\epsilon$~CMa, $\eta$~CMa \\
\hline
\end{tabular}
\end{center}
\end{table*}

The angular diameter of the bright southern supergiant
$\delta$~CMa (F8\,Iab)(HR 2693, HD 54065) has been measured with
the Narrabri Stellar Intensity Interferometer \citep{74hbdanda}
and twice with SUSI at 442\,nm \citep{99susi2}.  For this reason
it was chosen as an early observational target for the new red
system to provide an illustration of its performance.

The observations of $\delta$~CMa were made using the 700\,nm
filters (see Section~\ref{sec:efflam}) employing 5, 10, 20, 30 and
40\,m baselines. Each observation of $\delta$~CMa was bracketed by
observations of calibrator stars. The primary calibrators were
$\epsilon$~CMa and $\eta$~CMa and these were used on all twelve
nights. $\sigma^{2}$~CMa was also used as a calibrator on four
nights of observation.  The observational data were processed with
the SUSI pipeline with seeing corrections
(Section~\ref{sec:pipe}). Details of the observations and
seeing-corrected calibrated values of $V^{2}$ are listed in
Table~\ref{tab:obs}. Observations made in 2003 on four additional
nights to those listed in Table~\ref{tab:obs} have been omitted.
Unfortunately, for these early observations with the new system,
on three of the nights insufficient seeing data were recorded for
the data to be successfully processed.  The fourth night involving
observations at the 10\,m baseline, made under poor observing
conditions, resulted in a value of V$^2$ lying some four sigma
from a uniform-disk angular diameter fit to the other data and it
has been omitted.

The uncertainties in the values of $V^{2}$ have been determined by
scaling the standard errors in the mean (SEM), determined from the
N sets of scans for each night, by an appropriate value of
student's $t$ to take into account the fact that they have been
determined from a small number of independent measurements. The
confidence interval chosen was 68\%, equivalent to that of the
standard deviation for a Gaussian distribution.  The scaled
uncertainties ($\sigma_{V^{2}}$) listed in Table~\ref{tab:obs} are
the revised SEM values.

The angular diameter of the equivalent uniform disk
$\theta_{\mathrm{UD}}$ was found by fitting the expression
$|2J_{1}(x)/x|^{2}$, where $x = \pi
b\theta_{\mathrm{UD}}/\lambda_{\rm{eff}}$, to the $V^{2}$ data.
$b$ is the projected baseline and $\lambda_{\rm{eff}}$ the
effective wavelength of the observation. For $\delta$~CMa the
effective wavelength has been determined specifically for the star
as it is a reddened supergiant.  The flux distribution given by
\cite{87kiehling} for $\delta$~CMa has been used to calculate the
effective wavelength, following the same procedure described in
Section~\ref{sec:efflam}, to be 695.6$\pm$2.0\,nm.  A direct
measurement of the effective wavelength at a baseline of 5\,m gave
a value of 695.1\,nm.  This differs from the computed value by
only one quarter of the adopted uncertainty and, because of the
large scatter shown in Figure~\ref{fig:efflam} for measured values
for main-sequence stars, the computed value has been adopted for
the analysis.

The broad spectral bandwidth used for the observations will result
in bandwidth smearing \citep{02tandd} arising from the variation
in limb darkening as a function of wavelength across the band and
from non-linearity in the fitted curve.  An investigation of this
effect for $\delta$~CMa showed that the effect was negligible for
all baselines except 40\,m.  At 40\,m the combined effects of
limb-darkening and curvature increased the observed $V^{2}$ by
1.7\% relative to a monochromatic measurement at the effective
wavelength.  Therefore, the observed value of $V^{2}$ listed in
Table~\ref{tab:obs} has been reduced by this amount prior to the
fit for the uniform-disk angular diameter.

The uniform-disk angular diameter determined from the fit to the
values of $V^{2}$ in Table~\ref{tab:obs}, with the value at 40\,m
reduced to 0.47, is $3.457\pm0.019$\,mas.  It is noted that the
effect of the change in value of $V^{2}$ for 40\,m on the value
for $\theta_{\rm{UD}}$ is only 0.001\,mas. The fit also gives the
$V^{2}$ value at zero baseline equal to $1.003\pm0.009$. The
reduced $\chi^{2}$ for the fit is 1.73 indicating that the
uncertainties in the measured values of $V^{2}$ are underestimated
and this is believed to be due to incomplete correction of
residual seeing effects.  The uncertainties in the results of the
fit have therefore been increased by $\sqrt{1.73}$. The
uncertainty in the effective wavelength, given above, is
$\sim\pm0.3$\%.  Taking all uncertainties into account we find
$\theta_{\rm{UD}} = 3.457\pm0.024$\,mas ($\pm$0.7\%) and the zero
baseline $V^{2} = 1.003\pm0.011$.

\begin{figure}[h]
\begin{center}
\resizebox{\hsize}{!}{\includegraphics{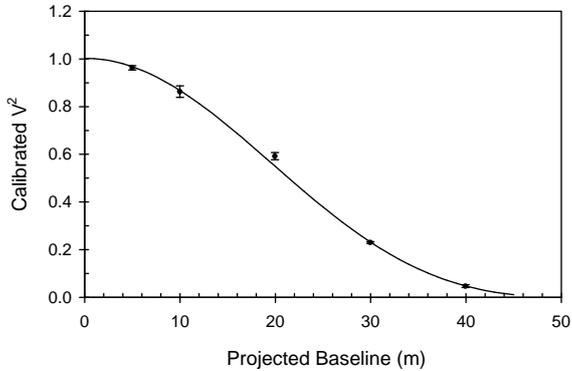}}
  \caption{Weighted mean values of calibrated $V^{2}$ versus
  projected baseline for $\delta$ CMa.  The curve is a least-squares
  fit to the observations for a uniform disk model (see text for details).}
  \label{fig:delcma}
  \end{center}
\end{figure}

Weighted mean values for the $V^{2}$ at each baseline are plotted
in Figure~\ref{fig:delcma} together with the fitted curve.

The new value for the uniform-disk angular diameter is listed in
Table~\ref{tab:dcma1} with the previous measurement made with the
Narrabri Stellar Intensity Interferometer \citep{74hbdanda} and
the two measurements made at 442\,nm with SUSI \citep{99susi2} .
The uncertainty in the angular diameter has been steadily reduced.
Note that, since the equivalent uniform-disk angular diameter is a
function of wavelength, the values in Table~\ref{tab:dcma1} cannot
be compared directly with one another.

\begin{table}[h]
\begin{center}
  \caption{The uniform-disk angular diameter of $\delta$~CMa determined with the NSII and with SUSI.}
  \label{tab:dcma1}
  \begin{tabular}{cccc}
\hline Instrument & $\lambda$ & $\theta_{\mathrm{UD}}$ &  $\sigma$\% \\
 & (nm) & (mas) &  \\
 \hline
 NSII & 443.0 & $3.29\pm0.46$   & 14.0 \\
 SUSI & 442.0 & $3.474\pm0.091$ &  2.6 \\
 SUSI & 442.0 & $3.535\pm0.090$ &  2.5 \\
 SUSI & 695.6 & $3.457\pm0.024$ &  0.8 \\
\hline
\end{tabular}
\end{center}
\end{table}

The true, limb-darkened angular diameter of $\delta$ CMa, for each
value of the uniform-disk angular diameters listed in
Table~\ref{tab:dcma1}, has been obtained using the appropriate
correction factor interpolated from the tabulation of
\citet{00dtb}.  For this purpose the following physical parameters
for $\delta$~CMa were adopted: effective temperature
$T_{\rm{eff}}=6000\pm200$\,K based on a number of values in the
literature; $\log{g}=0.6$ and [Fe/H]=0.19 from \citet{85landl}.
The effective temperature has been determined subsequently with
the aid of the new angular diameter to be 5818$\pm$53\,K
\citep{07delcma}.  The limb-darkening correction factors were
checked using the revised temperature with the same values for
$\log{g}$ and [Fe/H].  The only change was for 695.6\,nm with an
increase from 1.050 to 1.051.  Although this has negligible
effect, reducing the temperature by only 2\,K, the revised value
has been used.  The correction factors and the resulting values
for the limb-darkened angular diameter are listed in
Table~\ref{tab:dcma2}.  The accuracy of the correction factors
depends on how well the models represent the star but we note that
the interpolation in the tabulation by \citet{00dtb} has an
uncertainty of $\sim\pm0.002$. The limb-darkened angular diameter
values represent the same quantity and can be compared.

\begin{table}[h]
\begin{center}
  \caption{The limb-darkened angular diameter of $\delta$~CMa.  $\rho_{\lambda}$ is the ratio of limb-darkened
  to uniform-disk angular diameter from the tabulation by
  \citet{00dtb} used to convert the uniform-disk angular diameters
  in Table~\ref{tab:dcma1} to the limb-darkened angular diameters in
  this table.}
  \label{tab:dcma2}
  \begin{tabular}{cccc}
\hline Instrument & $\lambda$ & $\rho_{\lambda}$ & $\theta_{\mathrm{LD}}$ \\
 & (nm) &  & (mas) \\
 \hline
 NSII & 443.0 &  1.099 & $3.62\pm0.51$ \\
 SUSI & 442.0 &  1.100 & $3.82\pm0.11$ \\
 SUSI & 442.0 &  1.100 & $3.89\pm0.11$ \\
 SUSI & 695.6 &  1.051 & $3.633\pm0.026$ \\
\hline
\end{tabular}
\end{center}
\end{table}

The uncertainty in the NSII value for the limb-darkened angular
diameter is large and includes all three values determined with
SUSI.  However, the new 695.6\,nm SUSI value differs from each of
the 442\,nm values by more than the sum of the individual
uncertainties in each case.  Although the two 442\,nm values agree
with one another we believe that the new value is the most
reliable because of the improvement in calibration and seeing
correction techniques that have been developed for the red system.

The blue measurements were made during the commissioning phase of
SUSI and the analysis was based on observation and calibration
procedures which have since been significantly improved as a
result of experience.  For example, the current observing
procedure developed with the red system is to use at least two
calibrators, preferably positioned either side of the programme
star in the sky, to bracket all observations of the programme star
between observations of calibrators, and to minimise the cycle
time for a calibrator--programme star--calibrator sequence of
observations to minimise the effects of changing seeing
conditions.  At the shorter blue wavelengths, where both spatial
and temporal effects of seeing are greater, these changes in
procedure are even more important.  For the blue observations only
one calibrator was used at a time and not all observations of
$\delta$~CMa were bracketed by the calibrator. The cycle time for
a calibrator--programme star--calibrator sequence of observations
averaged 57 minutes compared with 18 minutes for the new red
measurement.

In view of these shortcomings the blue analysis was re-visited and
new fits to the data were made to determine the angular diameter.
Observations which were not bracketed by calibrator observations
were omitted and, in the case of observations calibrated by
$\epsilon$~CMa, this reduced the number of individual values of
$C$ from 127 to 94 and, in the case of $\eta$~CMa, from 94 to 88.
Fits were then made to the individual measurements of $V^{2}$,
rather than to weighted means for each baseline as was done in the
original analysis.  The resulting values for the limb-darkened
angular diameter of $\delta$~CMa are, for $\epsilon$~CMa as
calibrator, $3.75\pm0.11$\,mas and, for $\eta$~CMa as calibrator,
$3.70\pm0.17$\,mas.  While a discrepancy with the new value
remains, it is $\sim1.1\sigma$ in one case and $\sim0.4\sigma$ in
the other.  The agreement between the original blue values in
\citet{99susi2}, which differed significantly from the new value,
must be regarded as fortuitous and the re-examination of the blue
data supports the conclusion that the new value is the most
reliable because of the improvements in observation, calibration,
and seeing correction techniques that have been developed for the
red system.

A more detailed discussion of the results and the combination with
flux distributions to determine the emergent flux and effective
temperature for the star is in preparation \citep{07delcma}.

\subsection{Blue System Tests}\label{sec:blue}

A number of tests have been carried out with the red and blue
systems operating simultaneously to evaluate the performance of
the updated blue system.  These include the alignment of the beams
in the blue system, adjusting the optical paths differentially in
the blue system to ensure path equality at the beam-combiner,
calibrating the wavelength scale, and correcting the blue data for
differential optical path length variations and seeing effects
using data from the red system.  In each test the red system was
continuously scanning the fringe envelope and providing feedback
to the optical path length compensator to minimise the
differential path length variations. The blue data were recorded
and processed as discussed by \citet{99susi2}.

\subsubsection{Beam Alignment for the Blue
System}\label{sec:bluealign}

The tip-tilt correction system aligns and maintains alignment of
the beams for the red system but, for the blue system, the initial
alignment of the beams must be carried out separately.  Once
aligned, the tip-tilt system will maintain the alignment for the
blue system. The alignment is achieved using the original
reference detector for the blue system \citep{99susi1}.  The north
and south beams are aligned one at a time by closing shutters to
isolate the beam being aligned.  The beam is then aligned on the
reference detector, with the aid of either meters or an
oscilloscope displaying the vertical and horizontal positions of
the image, using the picomotor actuators on the back of the mirror
directing the beam into the beam-combiner.  Once aligned at the
start of an observing night it is a matter of less than
$\sim30$\,seconds to check the alignment during the course of the
night.  Experience has shown it to be extremely stable unless the
differential optical path length is adjusted by moving the north
input mirror on its slide (see Section~\ref{sec:delay}).

\subsubsection{Optical Path Matching for the Blue System}\label{sec:delay}

\begin{figure}
\begin{center}
\includegraphics[scale=0.53]{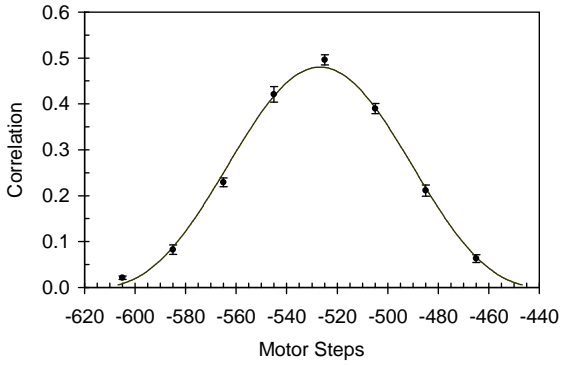}
\caption{An example of a delay curve measured with the blue system
for $\alpha$~Vir at a baseline of 5\,m.  The curve is a fit to the
raw measurements of $V^{2}$ and the position of the peak is
$-526.8\pm0.6$\,motor steps.  Further details are given in the
text.} \label{fig:delay}
\end{center}
\end{figure}

It was pointed out in Section~\ref{sec:blue} that matching the
paths for the red system does not necessarily do the same for the
blue system and there is likely to be a fixed optical path offset
between the beams in the blue system.  It was also noted that the
mirror directing the north beam to the blue beam-combiner is
mounted on a stepper motor driven slide.  In order to establish
the fixed optical path offset between the beams on the blue table,
and hence the matched path condition for the blue system, the
slide-mounted mirror was stepped through the fringe envelope to
produce a `delay curve', the peak of which occurs at the matched
path position. This was done with the red system maintaining
matched optical paths on the red table.  A typical delay curve is
shown in Figure~\ref{fig:delay} in which the measured $V^{2}$ is
plotted against the number of motor steps from the zero position
of the slide.  One motor step is equal to a differential change in
optical path length of 1.73$\mu$m.  The measurements were made at
a central wavelength of 480\,nm with the slit width set to give a
spectral bandwidth of 1.5\,nm.  The fitted curve has the form of
$|\sin{x}/x|^{2}$, corresponding to the near rectangular spectral
bandpass defined by the slits in the blue system.  We note that
the translation of the slide does not maintain the orientation of
the mirror precisely and small angular changes in the direction of
the reflected beam occur.  To overcome this the alignment of the
north beam was checked, as described in
Section~\ref{sec:bluealign}, whenever the mirror was translated by
$\geq20$\,steps.  The re-alignment takes no more than 30\,seconds
and the fit of the delay curve in Figure~\ref{fig:delay} and the
high value of $V^{2}$ for the peak of the curve are indicative of
the success of this approach.

\subsubsection{Calibration of the Wavelength Scale}

\begin{figure}
\centerline{%
\includegraphics[scale=0.55]{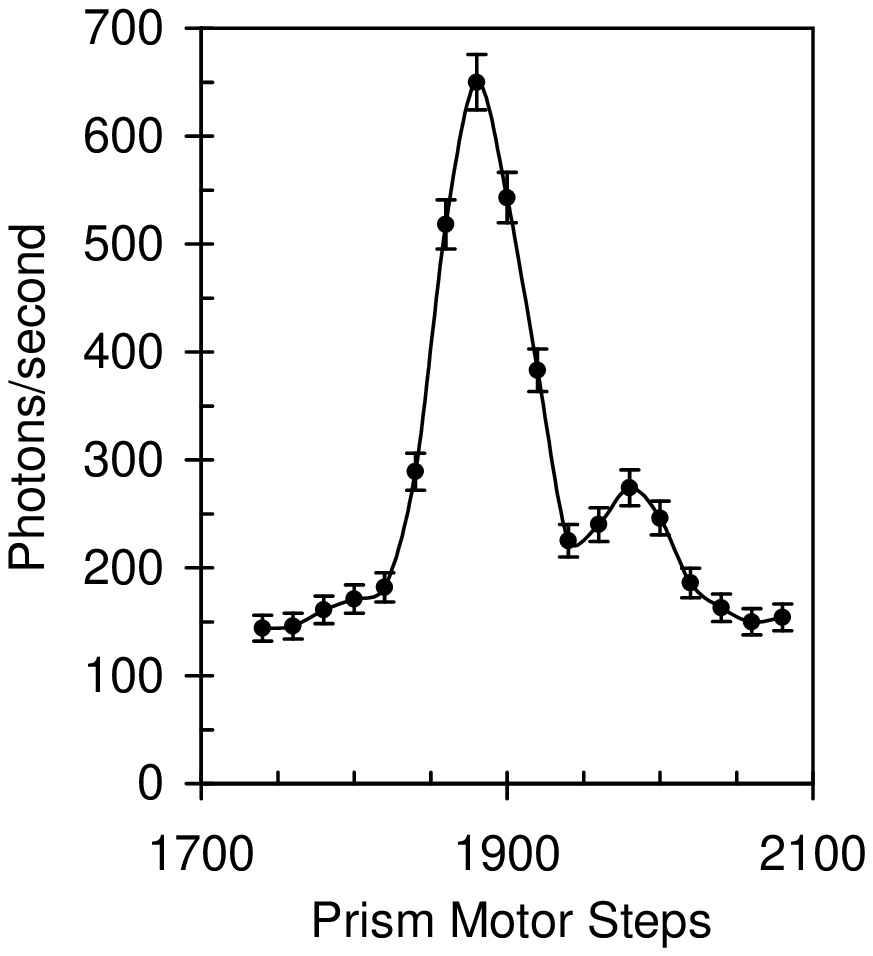}
\hspace{-4mm}
\includegraphics[scale=0.55]{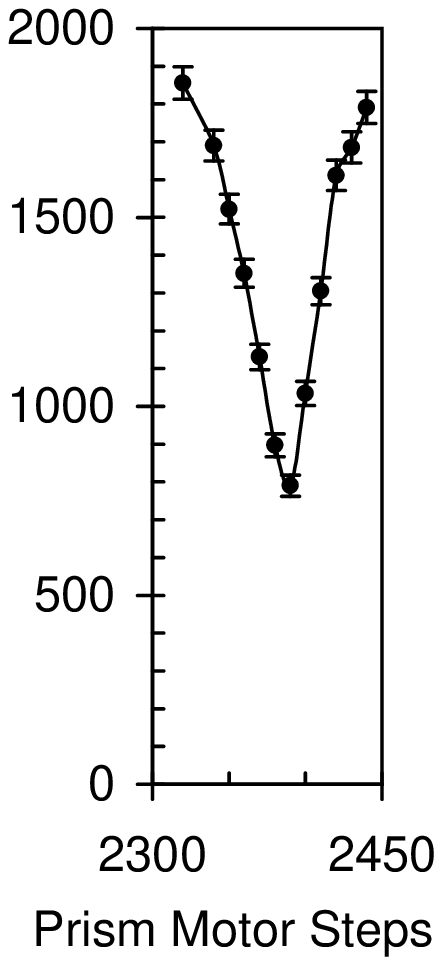}}
 \centerline{\hspace{18mm}(a)\hspace{28mm}(b)}
\caption{Scans through (a) the CIII-IV emission lines at
$\lambda$465\,nm for $\gamma^{2}$~Vel and (b) the core of the
H$\beta$ absorption line for $\alpha$~CMa using the rotating prism
mounts as described in the text.  In each case the spectrograph
slit was set to give a spectral bandpass 1\,nm wide, corresponding
to ~20 prism motor steps.} \label{fig:gamvel}
\end{figure}

The wavelength is selected by rotating the prism mount and the
general form of the relationship between wavelength and prism
motor steps has been calculated.  However, it needs to be
calibrated with observations of sources of known wavelength.  A
combination of laser lines and stellar emission and absorption
lines have been used and examples of stellar line observations are
shown in Figure~\ref{fig:gamvel}.  In each case a 30\,s
integration was made at each position as the prism was rotated a
few steps at a time.

\subsubsection{Correction of Blue System
Observations}\label{sec:bluecorr}

\begin{figure}
\begin{center}
\includegraphics[scale=0.54]{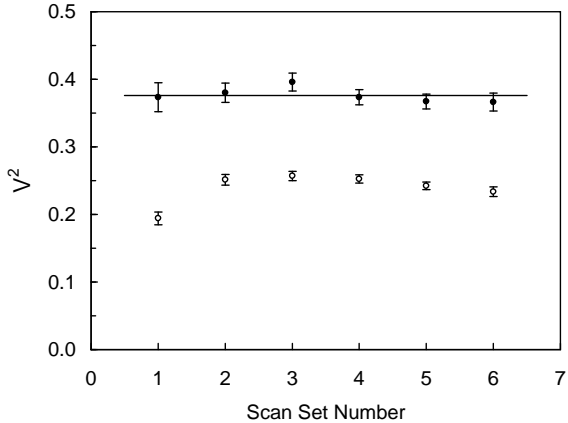}
\caption{An example of the effect of applying corrections for
differential optical path tracking errors and seeing effects.  The
data were obtained at a baseline of 20\,m for $\beta$~Cru. The
data points represented by open circles are the raw measurements
of $V^{2}$.  The filled circles represent the same data after
correction.  The horizontal line is the mean of the corrected data
points.  Further details are given in the text.}
\label{fig:corrdata}
\end{center}
\end{figure}

It was anticipated that optical path tracking errors and residual
seeing errors would reduce the value of $V^{2}$ measured with the
blue system. A number of observations have been made in which the
integration time for the blue system matched the time of
200\,seconds for a simultaneous set of scans with the red system.
The loss in $V^{2}$ due to optical path tracking errors, which
were derived from the recorded positions of the centres of the red
system fringe envelopes, was computed and a correction applied to
the measured blue $V^{2}$.  The seeing correction derived for the
red observations was also applied as a correction to the blue
$V^{2}$. The application of these two corrections significantly
improved the stability of the blue measurements in all cases.  An
example is shown in Figure~\ref{fig:corrdata}.  The observations
were made with a baseline of 20\,m and a spectral bandwidth of
4\,nm centred on a wavelength of 460\,nm.  The seeing was good but
variable and the improvement is clearly seen. The rms scatter of
the data has been reduced from $\pm9.8$\% to $\pm2.9$\%.

\subsection{Sensitivity Limits}\label{sec:sensy}

There are several factors that have contributed to an increase in
sensitivity of the red system compared to the original blue
system. These include the increased quantum efficiency of the APD
detectors compared to photomultipliers, the larger value of
$r_{0}$ at the longer wavelength ($r_{0}\propto\lambda^{6/5}$)
allowing the use of larger apertures, and the use of wider
spectral bandwidths.  The limiting sensitivity for a point source
at 700\,nm, which lies between the \textit{R} and \textit{I}
photometric bands, is $\sim$+5.  The gain in sensitivity relative
to the blue system operating at 442\,nm is a function of spectral
type and is summarised in Table~\ref{tab:comp}.

\begin{table}[h]
\begin{center}
  \caption{A summary of the visual magnitude gain in sensitivity, $\Delta$\textit{V}, of
  the red system compared to the original blue system as a function of spectral type.  \textit{V}$_{\mathrm{limit}}$ (a) is the limiting visual
  magnitude for observations with the red system's 700\,nm filter
  (\textit{V}$_{\mathrm{limit}}=+5$
  for spectral type A0) and \textit{V}$_{\mathrm{limit}}$ (b) is the limiting magnitude of the original blue
  system at 442\,nm (\textit{V}$_{\mathrm{limit}}=+2.5$ for spectral type A0).}
  \label{tab:comp}
  \begin{tabular}{cccc}
\hline Spectral & \textit{V}$_{\mathrm{limit}}$ & \textit{V}$_{\mathrm{limit}}$ & $\Delta$\textit{V} \\
Type  & (a) & (b) & \\
 \hline
 B0 & 4.7 & 2.8 & 1.9 \\
 A0 & 5.0 & 2.5 & 2.5 \\
 F0 & 5.3 & 2.2 & 3.1 \\
 G0 & 5.6 & 1.9 & 3.7 \\
 K0 & 5.8 & 1.7 & 4.1 \\
\hline
\end{tabular}
\end{center}
\end{table}

The sensitivity of the upgraded blue system has yet to be
established but a number of factors will contribute to its
improvement.  An increase in spectral bandwidth by up to a factor
of 2.5 may be possible.  The polarizing beam-splitters had
aberrations due to inhomogeneities and no significant increase in
signal-to-noise was obtained for effective aperture diameters on
the sky greater than $\sim$40\,mm.  Their removal allows larger
aperture diameters and both polarizations to be used.  It is
estimated that a gain in sensitivity between 1.5 and 3 magnitudes
will be obtained taking the limiting B magnitude to $>+$4.

\section{Summary}\label{sec:summary}

A beam-combination and detection system working at the red end of
the spectrum ($\lambda\lambda$500-950\,nm) has been installed and
commissioned in SUSI.  It has provided an increase in sensitivity
over the initial blue beam-combination and detection system that
ranges from $\sim$2 magnitudes at spectral type B0 to $>$4
magnitudes at K0.  It has also resulted in improved calibration
and accuracy of measurements of $V^{2}$ and hence improved
accuracy in the determination of angular diameters. Dichroic
beam-splitters have been introduced to allow simultaneous
observations to be made with red and blue beam-combination
systems.  The blue beam-combination system has been upgraded to
allow remote selection of wavelength in the range 430-500\,nm and
spectral bandpass in the range 1-10\,nm. Observations of
$\delta$~CMa made with the red system have been presented to
illustrate the improvements achieved and the results of various
tests on the two spectral systems have been discussed.

Results based on observations with the new red system include the
angular pulsations of the Cepheids $\ell$~Car and $\beta$~Dor
\citep{esowshop}, component mass and distance determinations for
the spectroscopic binaries $\lambda$~Sco \citep{lamsco} and
$\gamma^{2}$~Vel \citep{07north}, and dust scattering in Miras
resolved by optical interferometric polarimetry \citep{06dust}.
Both blue and red systems are now engaged in a programme of
observations of single, binary, pulsating, and rapidly rotating
stars.

\section*{Acknowledgments}
The SUSI programme has been funded jointly by the Australian
Research Council and the University of Sydney, with additional
support from the Pollock Memorial Fund and the Science Foundation
for Physics within the University of Sydney.  MJI acknowledges the
support of an Australian Postgraduate Award, JRN and APJ the
support of a University Postgraduate Award, and APJ and SMO the
support of Denison Postgraduate Awards.  JC was supported by MNRF
and EBS by ARC funds.  Collaboration with the CHARA group of the
Georgia State University and the exchange of ideas and the joint
development of software, particularly through the contributions of
T.A. ten Brummelaar, is acknowledged.



\begin{thebibliography}{}
\bibitem[Baldwin et al.(1994)Baldwin et al.]{coast}
  Baldwin, J.~E., Boyson, R.~C., Cox, G.~C., Haniff, C.~A.,
  Rogers, J., Warner, P.~J., Wilson, D.~M.~A., Mackay, C.~D. 1994,
  Proc. SPIE, 2200, 118
\bibitem[Bracewell(1965)Bracewell]{65rb}
  Bracewell R., 1965, The Fourier Transform and its Applications,
  McGraw-Hill, New York.
\bibitem[Davis et al.(2007a)Davis et al.]{07delcma}
  Davis, J., Booth, A.~J., Ireland, M.~J., Jacob, A.~P., North,
  J.~R., Owens, S.~M., Robertson, J.~G., Tango, W.~J., Tuthill,
  P.~G. 2007a, PASA, 24, in press, page number to come.
\bibitem[Davis et al.(2006)Davis et al.]{06spie}
  Davis, J., Ireland, M., Jacob, A.~P., North, J., Owens, S.~M.,
  Robertson, J.~G., Tango, W.~J. and Tuthill, P.~G. 2006, in Advances
  in Stellar Interferometry, ed. J.D. Monnier, M. Schöller and
  W.C. Danchi, Proc. SPIE, 6268, 626804
\bibitem[Davis et al.(2007b)Davis et al.]{esowshop}
  Davis, J., Ireland, M.~J., Jacob, A.~P., North, J.~R., Owens,
  S.~M., Robertson, J.~G., Tango, W.~J., Tuthill, P.~G. 2007b, in
  Proc. ESO Workshop ``The Power of Optical/IR Interferometry'', eds.
  F. Paresce \& A. Richichi, in press, Springer-Verlag, Berlin
\bibitem[Davis, Tango, \& Booth(2000)Davis et al.]{00dtb}
  Davis, J., Tango, W.~J., Booth, A.~J. 2000, MNRAS, 318, 387
\bibitem[Davis et al.(1999a)Davis et al.]{99susi1}
  Davis, J., Tango, W.~J., Booth, A.~J., ten Brummelaar, T.~A.,
  Minard, R.~A., Owens, S.~M. 1999a, MNRAS, 303, 77
\bibitem[Davis et al.(1999b)Davis et al.]{99susi2}
  Davis, J., Tango, W.~J., Booth, A.~J., Thorvaldson, E.~D.,
  Giovannis, J. 1999b, MNRAS, 303, 783
\bibitem[Hanbury Brown, Davis \& Allen(1974)Hanbury Brown et
  al.]{74hbdanda}
  Hanbury Brown, R., Davis, J., Allen L.~R. 1974, MNRAS, 167, 121
\bibitem[Ireland(2005)Ireland]{mjithesis}
  Ireland M. J. 2005, PhD thesis, University of Sydney
\bibitem[Ireland(2006)Ireland]{06mji}
  Ireland M. J. 2006, Proc. SPIE, 6268, 62680A-1
\bibitem[Ireland et al.(2006)Ireland et al.]{06dust}
  Ireland M.~J., Tuthill P.~G., Davis J., Tango W.~J. 2006,
  MNRAS, 361, 337
\bibitem[Kiehling (1987)]{87kiehling}
  Kiehling, R. 1987, A\&AS, 69, 465
\bibitem[Luck \& Lambert(1985)Luck \& Lambert]{85landl}
  Luck, R.~E., Lambert, D.~L. 1985, ApJ, 298, 782
\bibitem[North et al.(2007)North et al.]{07north}
  North, J.~R., Tuthill, P.~G., Tango, W.~J., Davis, J. 2007
  MNRAS, 377, 415
\bibitem[Seneta(1991)Seneta]{91ebs}
  Seneta E.~B., 1991, MSc thesis, University of Sydney
\bibitem[Silva \& Cornell(1992)Silva \& Cornell]{92silva}
  Silva, D.~R., Cornell, M.~E. 1992, ApJS, 81, 865
\bibitem[Tango \& Davis(2002)Tango \& Davis]{02tandd}
  Tango, W.~J., Davis, J. 2002, MNRAS, 333, 642
\bibitem[Tango \& Twiss(1980)Tango \& Twiss]{tandt}
  Tango, W.~J., Twiss, R.~Q. 1980, Progress in Optics, XVII, 239
\bibitem[Tango et al.(2006)Tango et al.]{lamsco}
  Tango W.~J. et al. 2006, MNRAS, 370, 884
\bibitem[Tuthill et al.(2004)Tuthill et al.]{04spie}
  Tuthill, P.~G., Davis, J., Ireland, M., North, J., O'Byrne, J.,
  Robertson, J.~G. and Tango, W.~J. 2004, in ``New Frontiers in Stellar Interferometry'',
  ed. W.~A. Traub, Proc. SPIE, 5491, 499
\end{thebibliography}
\end{document}